# Machine Learning for Building Energy and Indoor Environment: A Perspective

Zhijian Liu[1*], Di Wu[1], Hongyu Wei[1], Guoqing Cao[2]

[1]Department of Power Engineering, School of Energy, Power and Mechanical Engineering, North China Electric Power University, Baoding 071003, China; lzj6035@163.com (Z Liu); wudisjd@sina.com (D Wu); weihongyu1124@163.com (H Wei)

[2]Institute of Building Environment and Energy, China Academy of Building Research, Beijing 100013, PR China,cgq2000@126.com

Correspondence: lzj6035@163.com; Tel.: +86-0312-7522225; Fax: +86-0312-7522242

**Abstract:** Machine learning is a promising technique for many practical applications. In this perspective, we illustrate the development and application for machine learning. It is indicated that the theories and applications of machine learning method in the field of energy conservation and indoor environment are not mature, due to the difficulty of the determination for model structure with better prediction. In order to significantly contribute to the problems, we utilize the ANN model to predict the indoor culturable fungi concentration, which achieves the better accuracy and convenience. The proposal of hybrid method is further expand the application fields of machine learning method. Further, ANN model based on HTS was successfully applied for the optimization of building energy system. We hope that this novel method could capture more attention from investigators via our introduction and perspective, due to its potential development with accuracy and reliability. However, its feasibility in other fields needs to be promoted further.

## 1. Introduction

Building energy and environment are closely related to people's lifestyle. On one hand, how to reduce energy consumption has been considered as the prominent factor for economic growth, since the energy demand of residential and commercial buildings could reach 40% of total energy demand in both US and EU [1][2]. In China, this proportion is up to 30%, and 63% of that is utilized for heating and cooling space [3]. Also, energy consumption for building sectors account for approximately 30%-40% of the primary energy consumption in China [4]. On the other hand, the quality of indoor environment could severely impact the occupants' health. Aspects of building characteristics, occupant's lifestyle behaviors, air exchange rate as well as the relative humidity significantly associate to the indoor environment evaluation and the rate of respiratory-related diseases [5]. Therefore, the achievements for improving the built environment and refining the building structure are focus recently [6].

However, there still remain some serious obstacles for the improvement of building energy and environment: i) How to obtain the characteristics of a building energy system? ii) How to measure the indoor air quality? Technically, the most important puzzle is to find out the approaches that could perform rapid predictions and/or assessments, with good accuracy. The normal approaches are engineering method or statistical method, which are often used in many domains for performance evaluation. Engineering methods are comprehensive methods, which consists of several partial differential equations or needs to apply the physical principles, as well as thermal dynamics equations [7]. Fortunately, due to the dramatic advances of information technologies recently, many softwares provide the solution for the complex shape definition based on partial differential equations (PDE). Meanwhile, that promotes the development of Numerical Analysis of Dynamics in engineering practice. We could easily obtain the numerical solutions through software products, such as TRNSYS [8], ANSYS [9], FLUENT [10], COMSOL [11] etc. Engineering method based on large amounts of statistical data could achieve the highly accurate estimation [12–14]. Normally, engineering method could be regarded as the assumptions built on statistical information. Thus, we could try to develop an engineering model or equation through available statistical data for other regions [15–17]. However, there is a series of complex processes including the establishment of the models and equations, the setting of parameters and boundary conditions, which is difficult to satisfy the requirement that we aim to get the results instantly and conveniently. Especially in some unknown regions, we have no sophisticated theories for reference, which may become the obstacle for the development of engineering method. Statistical methods build the relationship between the target values and input variables through the speculated models based on the historical data. Statistical methods are suitable for the nonlinear relations and that avoid the defects of the engineering methods [18][19]. Nevertheless, large amounts of high-quality historical data require to be collected, and long time-consuming and large computer memory need to be considered. Besides, it will cause inaccurate and insignificant results, if the selection for analytical method is inappropriate.

Based on the above analyses, "Machine Learning" evolved from artificial intelligence [20]. The aim of machine learning method is to build one good predictor with no complicated process or other strict conditions. With the development of computer technology, machine learning method has been significantly developed and has been widely applied in more and more fields. Nevertheless, the utilization of machine learning method for energy conservation and indoor environment still remains in the infancy stage. Therefore, we expect that machine learning method could effectively contribute to more troublesome problems. In this article, we describe the possibility of the approach through several successful cases. Furthermore, we hope that this novel method could capture more attention from investigators via our introduction and perspective, and that the development of machine learning method could be further promoted.

## 2. Methodologies

### 2.1. Machine Learning Method

Machine learning method is a powerful technology that grew out of the exploration of artificial intelligence through the endeavor of scientists, which has huge potential for development [21–24]. Based on the historical data, machine learning method with suitable algorithms has the ability to "learn" the nonlinear relationship between the independent variables and target variables. Machine learning method focuses on the prediction through computers, and it is related to the computational statistics. Machine learning method started to flourish in the 1990s. It has been employed in many domains especially in the technologically advanced period, such as natural language processing, medical diagnosis, DNA sequencing, drug research etc. [25]. Robinson et al. [26] took advantage of machine learning method based on gradient boosting regression models to predict commercial building energy-consumption. Although this method was only successfully performed for building group in entire metropolitan areas, some limitations are existing through the experiments and verification. However, it is still indicated the universality of machine learning method. Sattlecker et al. [27] made use of machine learning method to recognize diagnostic spectral patterns in clinical practice, and emphasized the importance of the routine spectral data for the reanalysis process. Within the process of predict genome-wide with complex traits, González-Recio et al. [28] also suggest that machine learning method could achieve the prediction and classification, and deal with the multidimensional problems. Mateo et al. [29] verified that several machine learning methods and classical technologies could predict the short-term indoor temperature of buildings. They also indicated that the combination of Multilayer Perceptron (MLP) with non-linear autoregressive techniques (NARX) is the best predictor by comparison. In addition, Shirzadi et al. [30] and Pham et al. [31] verified that machine learning method is more accurate than statistical method for simulating volume of landslides (geology field). Moreover, they also indicated that support vector machines (SVM) model has the best prediction performances than other four models in this exploration. Therefore, we can see that machine learning method has already occupied the broad market.

However, for the domain of energy conservation and indoor environment, the theories of machine learning method are not sophisticated. It leads to the fact that this method has not been widely utilized in practice. The important reason is that the prediction structure with best performance is difficult to determine, when the different conditions was considered. Different models or algorithms selected and the number of input variables could substantially influence the accuracy and reliability of assessments. Thus, the selection of prediction structure undertaking the accurate estimators is extremely important. Machine learning method is mainly including Artificial Neural Networks (ANN), Support Vector Machines (SVM) etc. ANN as one epidemic model is utilized frequently by many investigators in different fields. It has strong ability to capture the complex nonlinear relationships between variables of input and output [32].

*2.2. Artificial Neural Networks (ANN)*

ANN structure is similar with the formation of "neurons", which could be divided into three parts, i.e. input layer, hidden layer and output layer (Figure 1) [33]. The training of ANN is in favor of acquirement for the suitable structure of model, which could avoid the risk of overfitting or underfitting. Many factors, e.g. historical data for training, input variables and different models, has the significant influence on the ANN structure. Accordingly, these aspects (e.g. the universality and representativeness for historical data for training, the appropriateness for model and input variables) should all be deliberated. In addition, based on various algorithms, different ANN structures could be achieved, such as multilayer feed-forward neural network (MLFNs) and general regression neural network (GRNN). Therefore, the aim for training is to determine the ANN structure including input variables, numbers of hidden nodes, hidden layer(s), and weight values.

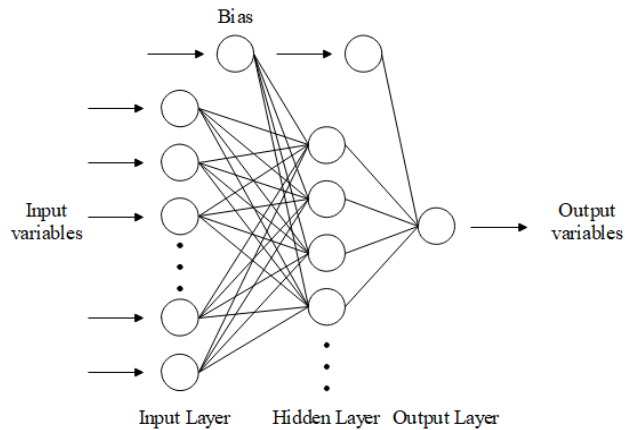

**Figure 1.** Schematic structure of ANN model [33].

*2.3. Application of ANN*

The measurement of indoor culturable fungi concentration is complicated and time-consumption, thus, Liu et al. [34] predicted the airborne culturable fungi concentration through machine learning method based on the data of indoor air quality.

For training and testing the structure of ANN, 249 groups of samples were measured. The measuring data include the input variables (indoor/outdoor PM2.5 & PM10, indoor temperature, relative humidity (RH) and $CO_2$ concentration) and input variable (concentration of indoor airborne culturable fungi) [35]. Through the comparison, it is indicated that ANN model has the good prediction performances than SVM [36]. Finally, ANN model with one hidden layer and 10 hidden nodes based on back-propagation algorithm was decided as the best structure. In the testing set, the prediction accuracy of the ANN model could reach 83.33% under the tolerance of ±30% (Figure 2) [34]. Therefore, the utilization of machine learning method for the measurement of indoor culturable fungi concentration provides a real-time estimation method with acceptable accuracy.

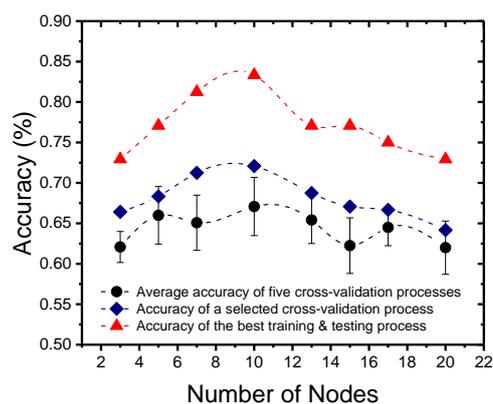

**Figure 2.** Trends of the accuracies vs number of nodes in the hidden layer of ANNs. Red points represent the accuracy of the best training and testing processes. Blue points represent the accuracy of a selected cross-validation process. Black points represent the average accuracy of five cross-validation processes. Reproduced with permission from Liu et al. [34].

In respect to the optimization process, it is no necessary to be known. That can be regarded as one "black-box" which can be realized by pro-process of training for model. It should be noted that regardless of weights or bias they have no physical significance. They only can be regarded as the mediums of the prediction process [37].

## 3. Development of Machine Learning Method

### 3.1. Hybrid Method Based on Machine Learning Method

According to the Gartner's 2016 hype cycle [38], machine learning method is at its peak of inflated expectations. Because of the complexity of objective and the huge amount of samples, the application of integrate machine learning method has developed promptly. Moss et al. [39] combine the discriminant analysis and machine learning method for enhancing the classification and prediction ability for a range of chemicals. Through the prediction process for solar radiation, Voyant et al. [40] suggest that hybrid method could give better results than single predictors. Marasco et al. [41] indicated that machine learning method (based on falling rule lists) is an rapid estimated tool for the potential of energy efficiency. Cramer et al. [42] proposed that machine learning method based on intelligent systems has the ability to predict rainfall. In the current era of big data, which instruments have the ability to classify the huge amount of data with quickly response speed will get the long-term development. Therefore, hybrid method based on machine learning method is regarded as the potential technology.

Liu et al. [43–48] confirmed that machine learning method (ANN model) based on high-throughput screening (HTS) is one effective approach to determine the building energy system, and it was successfully applied in the design of solar collector. HTS could screen huge amounts of combinations to obtain the candidates that satisfy the required properties, through high-throughput experimental together with computational techniques. It can promptly single out the promising candidates among millions of possible samples, which dramatically accelerated the progress for science [49]. In addition, HTS development is in the mature period, which is normally employed for the screening of the material with best properties. Due to the parallel and automatic disposition process, HTS has become one valuable selection technology [50].

### 3.2. Machine Learning Method Based on HTS

To integrate the advantages of above two technologies, machine learning method based on HTS was proposed [44]. Flow chart for this novel method is presented in Figure 3. Several extrinsic properties selected as the independent variables to input the first layer of ANN model. With the selection principle for extrinsic properties, independent variables should be easy-measured and are all relevant to the target [51]. The number of input variables could effect the accuracy of the method. Increasing the number of inputs will cause the complexity ulteriorly of models and prolong the time-consumption, while reducing the number will likely decrease the accuracy degree for prediction [52]. Thus, that should be determined by training and testing. Finally, the suitable predictions could be picked out based on HTS, and the verification could be carried out in practice further.

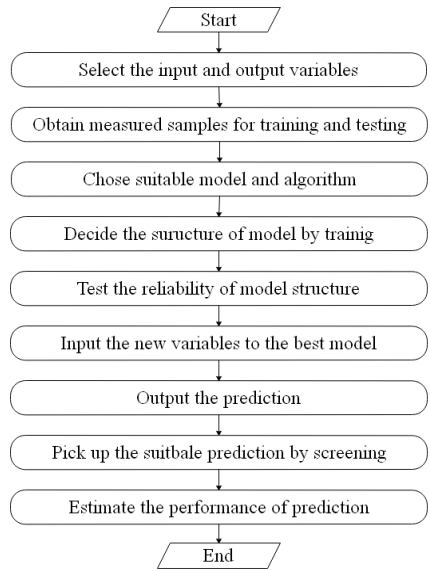

**Figure 3.** Flow chart for machine learning method based on HTS

*3.3. ANN based on HTS*

Due to the development of industrial production, solar collectors are wildly utilized in building for heating effect, which could effectively and reliably improve the indoor thermal environment [53]. However, the structure of solar collector with better performances is difficult to determine, when we consider various ambient conditions. Fortunately, Liu et al. integrate the ANN model and HTS to predict and optimize the structure of solar collector, and heat collection rates (HCR) was served as the target [43].

In this studied case, multilayer feed-forward neural network (MLFNs) were selected for model development, and back-propagation (BP) algorithm for training ANN based on the historical data. Sigmoid function was chosen as the transfer function. The best ANN configuration (learning rates=0.9, number of hidden nodes=7, number of training epochs=200 and momentums=0.9) was decided by comparison based on control variable method [44].

In order to explore the structure with better performance, 915 samples of extrinsic properties (i.e. tube length, tube number, tube center distance, tank volume, collector area, angle between tubes and ground) for solar collector were measured by portable test instruments, which is a simple assignment [47]. Through the prediction process based on HTS (Figure 4), two types of prediction with better HCR were selected whose characteristics are shown in Figure 4 (type 1 and type 2) [54]. This two structures have higher performance than previous one through the verification in practice, which indicates the feasibility of the novel integrated method.

In view of the complicated correlations between HCR and extrinsic properties, ANN model based on HTS could be considered as the efficient approach to predict and optimize the performance of solar collector.

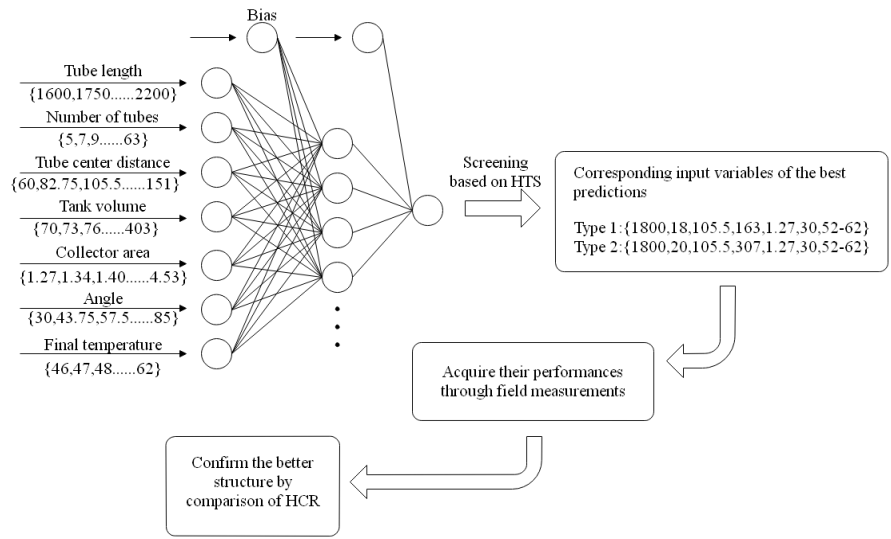

**Figure 4.** Flow chart for prediction process for solar collector by using ANN model based on HTS [44].

## 4. Discussion

The example about solar collector describes the successful application of building energy system design through machine learning method based on HTS with MLFNs to predict the better HCR. Simultaneously, Liu et al. also seek for the better solar collector structure with lower heat loss coefficient (HLC) [45]. They indicated that general regression neural network (GRNN) is the acceptable model to predict the HLC value. Therefore, the selection for model is various, which depends on the object studied. Whether the selection of model is suitable should be assessed through the training process.

Recently, several methods, such as Genetic Algorithm (GA), Ant Colony Optimization (ACO), Particle Swarm Optimization (PSO), have been successfully utilized for the optimization problems [55]. As efficient optimization techniques, they are specialized in solving complex problems with multivariable or nonlinear characteristics. They are great different from machine learning method in many aspects. For example, GA can imitate the process of biological evolution. Further, the next generations which have better properties could be reproduced from parent chromosomes by selection, crossover and mutation operation. The new offspring produces constantly until the termination condition is satisfied. Eventually, the last generations with better genes are the best solution for optimization problem. In addition, ACO is similar to the foraging behavior of ants. ACO takes advantage of ants' highly coordinated behavior to find out the solutions for optimal combinations. Through the update for pheromone and construction for path, the shortest path to the target locations could be discovered.

GA and ACO could achieve the global searching, but they do not calculate objective function of all possible combinations [56,57]. So their solutions may be trapped into local region, which results in the trouble that take the local optimal solution as the best result in the global area. Moreover, the value of characteristic parameters, such as number of population and generations, mutation rate, crossover rate (in GA) and evaporation ratio, error rate, number of ants (in ACO) has significant influence on the final results. However, the above parameters were determined only by experience without training process. Additionally, complicated programming also restricted their extensive application.

All combining forms of invariables have been taken into consideration by machine learning method based on HTS, which dramatically shorten the test period, avoids the defect of optimization methods and saves the operation time. Machine learning method has strong ties to mathematical optimization. Simultaneously, it delivers methods, models and applications to many fields. Nevertheless, almost all of references referred to machine leaning method or its optimization only indicate the availability in experiment [58][59]. Therefore, the suitable of that in other field remains to be verified so far, furthermore, its feasibility remain to be developed.

**5. Conclusion and Perspective**

Machine learning method has been employed in many fields with good accuracy and reliability. However, for the domain of energy conservation and indoor environment, its theories are not mature. In this paper, ANN model as one epidemic model was attempted to predict the indoor culturable fungi concentration and the structure of solar collector.

Through the utilization of ANN model, the relationship between air quality (i.e. indoor/outdoor PM2.5 & PM10, indoor temperature, RH and $CO_2$ concentration) and indoor culturable fungi concentration was built. Further, the prediction is achieved with instantaneity and reliability. In addition, the application of hybrid method has promoted the development of machine learning method up to a new stage. Machine leaning method based on HTS as a powerful prediction method was proposed recently, which could save the manpower and reduce time-consumption. Furthermore, AAN model based on HTS was successfully applied in the design for solar collector. It indicates that the novel method is an available approach in the design of building energy system.

Machine learning method drives from artificial intelligence. It focuses on prediction, based on known properties learned from the training data. Simultaneously, it also focuses on the models and methods from statistics and probability theory, whose goal is solve the problems of a practical nature [60]. Benefited from the development of digitized information, machine learning method has been promoted wildly both on the theoretical and practical aspects. However, the difficulty of development for effective machine learning is the difficulty of determining the model structure with higher reliability. Therefore, machine learning method often fail to deliver [61]. In addition, it is still need to be verified and discussed for the application of machine learning method in other fields to achieve excellent performance.

**Acknowledgments:** We are grateful to the Applied Sciences Editorial Office for the manuscript invitation. This work was supported by the National Key R & D Program of China-Source identification, monitoring and integrated control of indoor microbial contamination (No.2017YFC0702800), National Science Foundation of China (No. 51708211), and Natural Science Foundation of Hebei (No. E2017502051).

**Author Contributions:** Zhijian Liu and Hongyu Wei were responsible for the research about subject. Di Wu and Guoqing Cao were responsible for writing the manuscript.